\documentclass[reprint,pre,showpacs,amsmath,amssymb,aps]{revtex4-1}

\usepackage{natbib}
\usepackage{graphicx,color,rotating}
\usepackage[latin1]{inputenc}
\usepackage{textcomp}
\usepackage{dcolumn}
\usepackage{bm}     
\usepackage[version=3]{mhchem}  

\newcommand{\notop}{{{}_{}}}

\newcommand{\mr}[1]{\ensuremath{\mathrm{#1}}}
\renewcommand{\vec}[1]{\bm{#1}}

\newcommand{\pp}{\partial}       

\newcommand{\nablabf}{\boldsymbol{\nabla}}

\newcommand{\vJp}{\vec{J}_+}
\newcommand{\vJm}{\vec{J}_-}
\newcommand{\vJpm}{\vec{J}_{\pm}}

\newcommand{\vn}{\vec{n}}
\newcommand{\ver}{\vec{r}}

\newcommand{\cpm}{c_\pm}

\newcommand{\Dpm}{D^\notop_{\pm}}

\newcommand{\kB}{{k^\notop_\mathrm{B}}}

\newcommand{\epsw}{\epsilon_{\mathrm{w}}}

\newcommand{\bgam}{\bar{\gamma}}

\newcommand{\Gammax}{\Gamma_\mr{max}}

\newcommand{\lamc}{\lambda^{{}}_\mathrm{c}}
\newcommand{\lammax}{\lambda^{{}}_\mathrm{max}}
\newcommand{\lamD}{\lambda^{{}}_\mathrm{D}}

\newcommand{\lamDsqr}{\lambda^{{2}}_\mathrm{D}}

\newcommand{\VT}{V_{\mathrm{T}}}
\newcommand{\mupm}{{\mu_{\pm}^{{}}{}}}

\newcommand{\SIF}{\textrm{F}}

\newcommand{\SIJ}{\textrm{J}}

\newcommand{\SIK}{\textrm{K}}

\newcommand{\SIm}{\textrm{m}}

\newcommand{\SImM}{\textrm{mM}}

\newcommand{\SImum}{\textrm{\textmu{}m}}
\newcommand{\SInm}{\textrm{nm}}

\newcommand{\SIs}{\textrm{s}}

\newcommand{\beq}[1]{\begin{equation} \eqlab{#1}}
\newcommand{\eeq}{\end{equation}}
\newcommand{\bsub}{\begin{subequations}}
\newcommand{\esub}{\end{subequations}}
\def\bal#1\eal{\begin{align}#1\end{align}}
\def\bsubal#1\esubal{\bsub \begin{align}#1\end{align} \esub}
\newcommand{\nn}{\nonumber}
\newcommand{\eqlab}[1]{\label{eq:#1}}
\renewcommand{\eqref}[1]{Eq.~(\ref{eq:#1})}
\newcommand{\eqrefNoeq}[1]{(\ref{eq:#1})}

\newcommand{\figref}[1]{Fig.~\ref{fig:#1}}

\newcommand{\figlab}[1]{\label{fig:#1}}
\newcommand{\appref}[1]{Appendix~\ref{chap:#1}}

\newcommand{\chaplab}[1]{\label{chap:#1}}
\newcommand{\secref}[1]{Section~\ref{sec:#1}}

\newcommand{\seclab}[1]{\label{sec:#1}}
\newcommand{\tabref}[1]{Table~\ref{tab:#1}}

\newcommand{\tablab}[1]{\label{tab:#1}}

\begin{document}

\title{A sharp-interface model of electrodeposition and ramified growth}	

\author{Christoffer P. Nielsen}
\affiliation{Department of Physics, Technical University of Denmark, DTU Physics Building 309, DK-2800 Kongens Lyngby, Denmark}
\email{chnie@fysik.dtu.dk and bruus@fysik.dtu.dk}
\author{Henrik Bruus}
\affiliation{Department of Physics, Technical University of Denmark, DTU Physics Building 309, DK-2800 Kongens Lyngby, Denmark}

\date{26 June 2015, submitted to Phys.~Rev.~E}

\begin{abstract}

We present a sharp-interface model of two-dimensional ramified growth during quasi-steady electrodeposition. Our model differs from previous modeling methods in that it includes the important effects of extended space-charge regions and nonlinear electrode reactions. The model is validated by comparing its behavior in the initial stage with the predictions of a linear stability analysis.


\end{abstract}
\maketitle

\section{Introduction}

Electrodeposition is a technologically important process with diverse applications and implications, e.g. for battery technology, electroplating, and production of metal powders and microstructures \cite{Fleury2002,Rosso2007,Gallaway2014,Lin2011,Park2010,Scrosati2010, Tarascon2001, Winter2004, Han2014, Shin2003, Devos2007}. For well over a century it has, however, been known that the layer deposited during electrodeposition is prone to morphological instabilities, leading to ramified growth of the electrode surface. Over the years, a large number of experimental, theoretical, and numerical studies have been devoted to increasing the understanding of this ramified growth regime \cite{Chazalviel1990, Bazant1995,Sundstrom1995,Gonzalez2008,Nishikawa2013,Trigueros1991, Kahanda1989, Nikolic2006}. Big contributions to our understanding of the growth process have come from diffusion-limited aggregation (DLA) models \cite{Witten1983, Witten1981a} and, more recently, phase-field models similar to those which have successfully been applied to solidification problems \cite{Guyer2004, Guyer2004a, Shibuta2007, Liang2014, Cogswell2015}. However, while both of these approaches capture parts of the essential behaviour of ramified growth, they have some fundamental shortcomings when applied to the electrodeposition problem.

The first of these shortcomings has to do with the ion transport in the system. Typically, the electrolyte contains a cation of the electrode metal which can both deposit on the electrodes and be emitted from the electrodes. The anion, on the other hand, is blocked by the electrodes. The electrodes thus act as ion-selective elements, and for this reason the system exhibits concentration polarization when a voltage is applied. In 1967, Smyrl and Newman showed \cite{Smyrl1967} that in systems exhibiting concentration polarization, the linear ambipolar diffusion equation breaks down when the applied voltage exceeds a few thermal voltages. At higher voltages a non-equilibrium extended space-charge region develops next to the cathode, causing the transport properties of the system to change dramatically. It seems apparent that this change in transport properties must also lead to a change in electrode growth behavior. Indeed, this point was argued by Chazalviel already in his 1990 paper \cite{Chazalviel1990}. Now, the issue with DLA and phase-field models is that neither of these methods account for non-zero space-charge densities. It is therefore only reasonable to apply these methods in the linear regime, where the applied voltage is smaller than a few thermal voltages.

The other shortcoming of DLA and phase-field methods is their treatment of the electrode-electrolyte interface. It is well known in electrochemistry that electrodeposition occurs with a certain reaction rate, which is dependent on the electrode overpotential and typically modelled using a Butler--Volmer type expression \cite{Dukovic1990,Bazant2013}. Nevertheless, electrode reactions are neither included in DLA methods nor in phase-field methods.

There have been attempts to include finite space-charge densities in phase-field models, but the resulting models are only practical for 1D systems because they require an extremely dense meshing of the computational domain \cite{Guyer2004, Guyer2004a}. Attempts at including electrode reactions suffer from similar problems, as the proposed models are sensitive to the width of the interface region and to the interpolation function used in the interface region \cite{Liang2012, Liang2014}.

To circumvent the shortcomings of the established models we pursue a different solution strategy in this paper. Rather than defining the interface via a smoothly varying time-dependent parameter as in the phase-field models, we employ a sharp-interface model, in which the interface is moved for each discrete time step. Using a sharp-interface model has the distinct advantage that electrode reactions are easily implemented as boundary conditions. Likewise, it is fairly straight-forward to account for non-zero space charge densities in a sharp-interface model, see for instance our previous work Refs.~\cite{Nielsen2014,Nielsen2014b}.

Like most previous models, our sharp-interface model of electrodeposition models the electrode growth in two dimensions. There have been some experiments in which ramified growth is confined to a single plane and is effectually two dimensional \cite{Trigueros1991, Fleury1996, Leger1999,Leger2000}. However, for most systems ramified growth occurs in all three dimensions. There will obviously be some discrepancy between our 2D results and the 3D reality, but we are hopeful that our 2D model does in fact capture much of the essential behavior.

At this stage, our sharp-interface model is only applicable once the initial transients in the concentration distribution have died out. In its current form the model is therefore mainly suitable for small systems, in which the diffusive time scale is reasonably small. We aim at removing this limitation in future work.

\section{Model system}

The model system consists of two initially flat parallel metal electrodes of width $W$ placed a distance of $2L$ apart. In the space between the electrodes is a binary symmetric electrolyte of concentration $c_0$, in which the cation is identical to the electrode material. The electrodes can thus act as both sources and sinks for the cation, whereas the anion can neither enter nor leave the system. A voltage difference $V_0$ (in units of the thermal voltage $\VT=\kB T/e$) is applied between the two electrodes, driving cations towards the top electrode and anions toward the bottom electrode. A sketch of the system is shown in \figref{Geometry_sketch}.

 By depositing onto the top electrode we ensure that the ion concentration increases from top to bottom, so we do not have to take the possibility of gravitational convection into account. To limit the complexity of the treatment, we also disregard any electroosmotic motion, which may arise in the system. We note, however, that the sharp-interface model would be well suited to investigate the effects of electroosmosis, since the space charge density is an integral part of the model.

\begin{figure}[!t]
    \includegraphics[width=0.85\columnwidth]{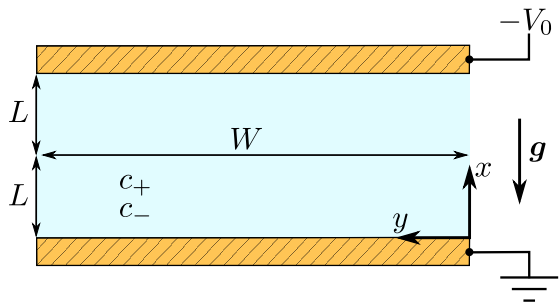}
    \caption{\figlab{Geometry_sketch} Sketch of the initial geometry of the system. Two co-planar metal electrodes of width $W$ are placed a distance of $2L$ apart. The gap between them is filled by an electrolyte with cation concentration $c_+$ and anion concentration $c_-$. A voltage difference of $V_0$ is applied between the electrodes. }
\end{figure}

\section{Solution method}
\seclab{Sol_method}

The basic idea in our solution method is to solve the transport-reaction problem for each time step, and then use the calculated currents to find the amount of material deposited at the electrode. Based on this deposition rate the geometry is updated, and the transport-reaction problem is solved for a new time step, as illustrated in \figref{Growth_sketch}.

\begin{figure}[!t]
    \includegraphics[width=0.7\columnwidth]{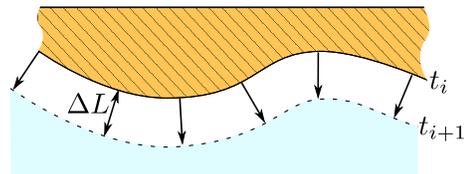}
    \caption{\figlab{Growth_sketch} Sketch of the electrode growth. The electrode surface at time $t_i$ is indicated with a full line. In the time step $t_{i+1}-t_i$ an amount of material $\Delta L$ is deposited on the electrode. On basis of the deposited material the geometry at time $t_{i+1}$ is created (indicated with a dashed line).}
\end{figure}

The major difficulty in employing this method is that when the geometry is updated the computational domain is also remeshed, so there is no straight-forward way of continuing from the old solution of the transport-reaction problem. One way of getting around this issue is to separate the time scales in the problem. More to the point, we assume that the growth of the electrode happens so slowly, compared to the transport time scales, that the transport problem always is in quasi steady-state. By treating the transport-reaction problem as being in steady state in each time step, a solution can be computed without reference to solutions at previous time steps.

Obviously, the quasi steady-state assumption is flawed in the initial time after a voltage is applied to the system, as the application of a voltage gives rise to some transients in the transport problem. However, after the initial transients have died out the assumption is quite reasonable, except for the case of very concentrated electrolytes. To see that, we consider the thickness $\Delta L$ of the electrode growth in a time interval $\Delta t$,
\begin{align}
  \Delta L = a^3 \Delta t J_+,    \eqlab{Simple_deltaL}
  \end{align}
where $a^3$ is the volume of a metal atom in the solid phase and $J_+$ is the current density of metal ions entering the electrode. The current density is on the order of the limiting current $2c_0D_+/L$, so the time scale associated with an electrode growth of $\Delta L$ is
\begin{align}
\Delta t = \frac{\Delta L}{a^3 J_+} \sim \frac{L\Delta L }{2D_+ c_0 a^3}.
\end{align}
On the other hand, the transport time scale $t_\mr{diff}^{\Delta L}$ associated with the distance $\Delta L$ is
\begin{align}
t_\mr{diff}^{\Delta L} \sim \frac{\Delta L^2}{2D_+}.
\end{align}
The ratio of the transport time scale to the growth time scale is thus
\begin{align}
\frac{t_\mr{diff}^{\Delta L}}{\Delta t} \sim \frac{\Delta L}{L} c_0 a^3,
\end{align}
which is indeed very much smaller than unity.

As mentioned above, our model does not apply to the initial time after the voltage is applied. To estimate how this impacts our results, we make a comparison of the important time scales. The time it takes for the transients to die out is given by the diffusion time,
\begin{align}
t_\mr{diff}^L = \frac{L^2}{2D_+}.
\end{align}
The growth rate of the most unstable harmonic perturbation to the electrode surface we denote $\Gamma_\mr{max}$ (see Ref.~\cite{Nielsen2015Arxiv}), and from this we obtain an instability time scale,
\begin{align}
t_\mr{inst} \sim \frac{1}{\Gamma_\mr{max}}.
\end{align}
It is apparent that if
\begin{align}
t_\mr{diff}^L \lesssim t_\mr{inst},
\end{align}
then nothing interesting happens to the electrode surface in the time it takes the transients to disappear. In this case our quasi-steady approach is therefore justified.

Even if $t_\mr{diff}^L \gg t_\mr{inst}$ our approach may be justified. If the total deposition time is much larger than $t_\mr{diff}^L$, then what happens in the time before the transients die out is largely unimportant for the growth patterns observed in the end. Thus, though the quasi-steady assumption seems restrictive, it actually allows us to treat a fairly broad range of systems.



\section{Governing equations}

\subsection{Bulk equations}

The ion-current densities in the system are given as
\bsubal
\vJpm &= - \Dpm c_0 \cpm \nablabf \mupm,   \eqlab{Currents} \\
\mupm &= \ln(\cpm) + z_\pm \phi, \eqlab{Chem_pot}
\esubal
where $\Dpm$ are the diffusivities of either ion, $c_0$ is the initial ion concentration, $\cpm$ are the concentrations of either ion normalized by $c_0$, $\mupm$ are the electrochemical potentials normalized by the thermal energy $\kB T$, and $\phi$ is the electrostatic potential normalized by the thermal voltage $\VT = \kB T/e$. In steady state the Nernst--Planck equations take the form
\begin{align}
0  = - \nablabf \cdot \vJpm.  \eqlab{NP}
\end{align}
The electrostatic part of the problem is governed by the Poisson equation,
\begin{align}
2 \lamDsqr \nabla^2 \phi  = -\rho =  -z_+c_+-z_-c_-,  \eqlab{Poisson}
\end{align}
where the Debye length $\lamD$ is given as
\begin{align}
\lamD = \sqrt{\frac{\kB T \epsw}{2e^2c_0}}.
\end{align}
At the electrodes the anion flux vanishes,
\begin{align}
\vn \cdot \vJm = 0,   \eqlab{No_flux}
\end{align}
and the cation flux is given by a reaction expression
\begin{align}
\vn \cdot \vJp = -R.   \eqlab{R_flux}
\end{align}
Rather than explicitly modelling the quasi-equilibrium Debye layers at the electrodes, we follow Ref.~\cite{Nielsen2014b} and implement a condition of vanishing cation gradient at the cathode,
\begin{align}
\vn \cdot \nablabf c_+ = 0.   \eqlab{Grad0}
\end{align}
The last degree of freedom is removed by requiring global conservation of anions,
\begin{align}
\int_\Omega \big(c_- -1 \big) \ \mr{d}V =0.   \eqlab{cm_cons}
\end{align}

\subsection{Reaction expression}

We model the reaction rate using the standard Butler--Volmer expression \cite{Sundstrom1995},
\begin{align}
R=k_0  \left [ c_+ e^{-\bgam \kappa+\alpha Z(\phi+V)}-e^{-\bgam \kappa -   (1-\alpha)Z(\phi+V) }\right ],   \eqlab{R}
\end{align}
where $k_0$ is the rate constant of the reaction, $V$ is the non-dimensionalized electrode potential, $\kappa$ is the surface curvature, $\alpha$ is the charge-transfer coefficient, and $\bgam$ is given in terms of the surface energy $\gamma$,
\begin{align}
\bgam = \frac{a^3 \gamma}{\kB T }.
\end{align}
Here, $a^3$ is the volume occupied by one atom in the solid phase. $\bgam \kappa$ is thus a measure of the energy per atom relative to the thermal energy.

\section{Numerical stability}

Due to the surface energy term in the reaction expression, the surface is prone to numerical instability. In an attempt to reach the energetically favorable surface shape, the solver will sequentially overshoot and undershoot the correct solution.
The fundamental issue we are facing is that the problem at hand is numerically stiff. As long as we are using an explicit time-integration method we are therefore likely to encounter numerical instabilities.

The straight-forward way of updating the position $\ver$ of the interface is to use the explicit Euler method,
\begin{align} \eqlab{Expl_Euler}
\ver(t+\Delta t) = \ver(t) + \vn a^3\Delta t R(t),
\end{align}
where $R(t)$ is the (position dependent) reaction rate at time $t$.
To avoid numerical instabilities, we should instead use the implicit Euler method,
\begin{align} \eqlab{Impl_Euler}
\ver(t+\Delta t) = \ver(t) + \vn a^3\Delta t R(t+\Delta t),
\end{align}
where the reaction rate is evaluated at the endpoint instead of at the initial point.
This is however easier said than done. $R(t+\Delta t)$ depends on $\ver(t+\Delta t)$ as well as on the concentration and potential distribution at $t+ \Delta t$. Even worse, through the curvature $R(t+\Delta t)$ also depends on the spatial derivatives of $\ver(t+\Delta t)$.

 The way forward is to exploit that only part of the physics give rise to numerical instabilities. It is therefore sufficient to evaluate the problematic surface energy at $t+\Delta t$ and evaluate the remaining terms at $t$. For our purposes we can therefore make the approximation
 \begin{align}
 R(t+\Delta t) \approx R\big(t,\kappa(t+\Delta t)\big),
 \end{align}
where $\kappa$ is the curvature. This does still make for a quite complicated nonlinear PDE, but we are getting closer to something tractable. The difference in curvature between $t$ and $t+\Delta t$ is small (otherwise we are taking too big time steps), so we can approximate
\begin{align}
R\big(t,\kappa(t+\Delta t)\big) \approx R\big(t,\kappa(t)\big) + R'\big(t,\kappa(t)\big)\Delta \kappa,
\end{align}
where $R'$ denotes $R$ differentiated with respect to $\kappa$ and $\Delta \kappa = \kappa(t+\Delta t)-\kappa (t)$.
The curvature can be written as
\begin{align}
\kappa = \frac{\pp \theta}{\pp s},
\end{align}
where $\theta$ is the tangential angle of the interface and $s$ is the arc length along the interface. We therefore have
\begin{align}
\Delta \kappa = \kappa(t+\Delta t)-\kappa (t) = \frac{\pp \theta_2}{\pp s_2}-\frac{\pp \theta_1}{\pp s_1},
\end{align}
where we have adopted the shorthand notation $1$ and $2$ for time $t$ and $t+\Delta t$, respectively. The arc lengths $s_1$ and $s_2$ will obviously differ for any nonzero displacement, but this is a small effect compared to the angle difference. As an approximation we therefore use $s_2 \approx s_1 $ and obtain
\begin{align}
\Delta \kappa \approx  \frac{\pp (\theta_2-\theta_1)}{\pp s_1}.
\end{align}
The tangential angle is a function of the surface parametrization,
\begin{align}
\tan(\theta_1) = \frac{\pp y_1}{\pp x_1}.
\end{align}
For small displacements we can approximate
\begin{align}
\tan(\theta_2)&= \frac{\pp y_2}{\pp x_2} = \frac{\pp (y_1+\Delta y)}{\pp (x_1+\Delta x)} \nn \\
&= \left (1+ \frac{\pp \Delta x}{\pp x_1 }  \right )^{-1}\left (\frac{\pp y_1}{\pp x_1} + \frac{\pp \Delta y}{\pp x_1} \right )\nn \\
&\approx \frac{\pp y_1}{\pp x_1} + \frac{\pp \Delta y}{\pp x_1} - \frac{\pp y_1}{\pp x_1}\frac{\pp \Delta x}{\pp x_1 } \nn\\
&= \tan(\theta_1) + \frac{\pp \Delta y}{\pp x_1} - \tan(\theta_1)\frac{\pp \Delta x}{\pp x_1 }.
\end{align}
The difference in tangential angles can then be written
\begin{align}
\theta_2 - \theta_1 &= \arctan\left [\tan(\theta_1) + \frac{\pp \Delta y}{\pp x_1} - \tan(\theta_1)\frac{\pp \Delta x}{\pp x_1 }\right ] - \theta_1 \nn \\
&\approx   \frac{1}{1+ \tan^2(\theta_1)} \bigg[\frac{\pp \Delta y}{\pp x_1} - \tan(\theta_1)\frac{\pp \Delta x}{\pp x_1 }\bigg].   \eqlab{Delta_theta}
\end{align}
Returning to the implicit Euler method \eqref{Impl_Euler}, we project it onto the normal vector to obtain
\begin{align}
\Delta L &= a^3\Delta t R(t+\Delta t) \nn \\
&\approx a^3\Delta t\left [R\big(t,\kappa(t)\big) + R'\big(t,\kappa(t)\big)\Delta \kappa\right ], \eqlab{DL_impl_Euler}
\end{align}
where $\Delta L = \vn \cdot \big[ \ver(t+\Delta t) - \ver(t) \big]$. The increments in the $x$ and $y$ directions are related to $\Delta L$ via
\begin{align}
\Delta x = n_x \Delta L, \quad \Delta y = n_y \Delta L.
\end{align}
Inserting these in \eqref{Delta_theta} and writing out the curvature difference $\Delta \kappa$, we obtain a linear PDE for the displacement $\Delta L$
\begin{align}  \eqlab{Lin_PDE_impl_Euler}
& \frac{\Delta L - a^3\Delta t R\big(t,\kappa(t)\big) }{a^3\Delta t R'\big(t,\kappa(t)\big)}  \nn \\ & \qquad =\Delta \kappa
 = \frac{\pp }{\pp s_1}\left \{   \frac{ n_y - n_x\tan(\theta_1)}{1+ \tan^2(\theta_1)}\frac{\pp \Delta L}{\pp x_1} \right \}.
\end{align}
In the limit $\Delta  \kappa =0$ this equation reduces to the original forward Euler method \eqrefNoeq{Expl_Euler}.

\subsection{Correction for the curvature}

In the previous derivation, we did not take into account that the local curvature slightly changes the relation between amount of deposited material and surface displacement $\Delta L$. The deposited area in an angle segment $d \theta$ can be calculated as
\begin{align}
 dA &=  \frac{d\theta}{2\pi}\left [ \pi \left (\frac{1}{\kappa} + \Delta L\right )^2 - \pi \frac{1}{\kappa^2}\right ] \nn \\
 &=\frac{d\theta}{2}\left [ \Delta L^2 +2 \frac{\Delta L}{\kappa}\right ].
\end{align}
The line segment $ds$ is related to the angle segment as $ds = d\theta/\kappa$. This means that
\begin{align}
a^3 \Delta t R(t+\Delta t)&=\frac{dA}{ds} = \frac{\kappa}{2}\left [ \Delta L^2 +2 \frac{\Delta L}{\kappa}\right ]\nn \\
& = \Delta L + \frac{\kappa}{2} \Delta L^2.
\end{align}
Using this expression in \eqref{DL_impl_Euler} yields the slightly nonlinear PDE, with the term $\frac{1}{2}\kappa \Delta L^2$,
\begin{align}  \eqlab{NonLin_PDE_impl_Euler}
& \frac{\Delta L +\frac{\kappa}{2} \Delta L^2- a^3\Delta t R\big(t,\kappa(t)\big) }{a^3\Delta t R'\big(t,\kappa(t)\big)} \nn \\
  &\qquad \qquad \qquad  =\frac{\pp }{\pp s_1}\left \{   \frac{ n_y - n_x\tan(\theta_1)}{1+ \tan^2(\theta_1)}\frac{\pp \Delta L}{\pp x_1} \right \},
\end{align}
in place of \eqref{Lin_PDE_impl_Euler}.

\section{Noise}
\seclab{Noise}

\begin{figure}[!t]
    \includegraphics[width=0.9\columnwidth]{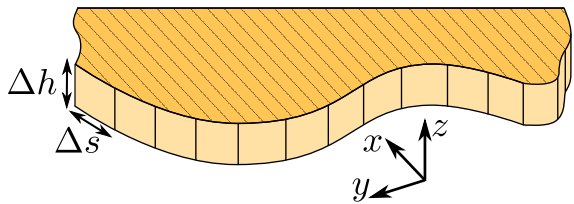}
    \caption{\figlab{3D_electrode_sketch} (Color online) Three-dimensional extension of our two-dimensional model. The electrode interface can vary in the $xy$-plane according to the calculated ion-currents, but it has a fixed depth $\Delta h$ in the $z$-direction. The interface is also divided into a number of bins of width $\Delta s$ in the $xy$-plane. Each bin thus has the area $\Delta h \Delta s$.   }
\end{figure}

An important part of the problem is the noise in the system, since the noise is what triggers the morphological instability and leads to formation of dendrites. Exactly how the noise should be defined is however a matter of some uncertainty. Most previous work uses a thermal white noise term with a small, but seemingly arbitrary amplitude. In this work we use a slightly different approach, in which we assume that the noise is entirely attributed to shot noise.

As it turns out, this approach requires us to be more specific about how our 2D model is related to the three-dimensional reality. In \figref{3D_electrode_sketch} a sketch of the tree-dimensional electrode is shown. The electrode interface is free to vary in the $xy$-plane, but has a fixed depth $\Delta h$ in the $z$-direction. Obviously, most real electrodeposits will have a more complicated behavior in the $z$-direction, but for electrodeposits grown in a planar confined geometry this is actually a reasonable description.

Solving the transport-reaction problem yields the current density at each point along the electrode surface, that is the average number of ions arriving per surface area per time. The mean number $Q$ of ions arriving in an electrode section of size $\Delta h \Delta s$ in a time interval $\Delta t$ is thus
\begin{align} \eqlab{N_avg}
Q = J_+ \Delta h \Delta s \Delta t.
\end{align}
Since the ions are discrete entities, the actual number of arriving ions will, however,  fluctuate randomly around the mean $Q$ with some spread $\sigma$. We assume that within the time interval $\Delta t$, the arrival of each ion is statistically uncorrelated with the arrival of each other ion. It can then be shown that, as long as $Q\gtrsim 10$, the number of arriving ions follow a normal distribution with mean $Q$ and standard deviation
\begin{align}
\sigma = \sqrt{Q}.
\end{align}
This corresponds to an extra random current density
\begin{align}
J_\mr{rand} = \frac{\sqrt{Q}}{\Delta h\Delta s \Delta t} q_\mr{rand} = \sqrt{ \frac{J_+}{\Delta h\Delta s \Delta t}} q_\mr{rand},
\end{align}
where $q_\mr{rand}$ is a random number taken from a normal distribution with mean 0 and standard deviation 1. This in turn corresponds to a random electrode growth of
\begin{align}  \eqlab{DL_rand}
\Delta L_\mr{rand} = a^3 \sqrt{\frac{J_+ \Delta t}{\Delta h\Delta s}} q_\mr{rand}.
\end{align}
Now, there is something slightly weird about this expression for the random growth: it seems that the random growth becomes larger the smaller the bin size $\Delta s$ is. However, as the bin size becomes smaller the weight of that bin in the overall behavior is also reduced. The net effect is that the bin size $\Delta s$ does not matter for the random growth, see \appref{Init_growth} for a more thorough treatment.

The bin depth $\Delta h$, on the other hand, does matter for the random growth. Since our model is not concerned with what happens in the $z$-direction, we simply have to choose a physically reasonable value of $\Delta h$, and accept that our choice will have some impact on the simulations. This is a price we pay for applying a 2D model to a 3D phenomenon.


\section{Numerical solution}

To solve the electrodeposition problem we use the commercially available finite element software \textsc{COMSOL Multiphysics} ver.~4.3a together with \textsc{MATLAB} ver.~2013b. Following our previous work~\cite{Nielsen2014,Nielsen2014b,Gregersen2009}, the governing equations and boundary conditions Eqs.~\eqrefNoeq{Currents}, \eqrefNoeq{Chem_pot}, \eqrefNoeq{NP}, \eqrefNoeq{Poisson}, \eqrefNoeq{No_flux}, \eqrefNoeq{R_flux}, \eqrefNoeq{Grad0}, \eqrefNoeq{cm_cons}, \eqrefNoeq{R}, and \eqrefNoeq{NonLin_PDE_impl_Euler} are rewritten in weak form and implemented in the mathematics module of \textsc{COMSOL}. For each time step the following steps are carried out: First, a list of points defining the current electrode surface is loaded into \textsc{COMSOL}, and the surface is created using a cubic spline interpolation between the given points. The computational domain is meshed using a mesh size of $\Delta s$ at the electrode surface, a mesh size of $l$ in a small region next to the electrode, and a much coarser mesh in the remainder of the domain. Next, the curvature of the surface is calculated at each point. The solution from the previous time step is then interpolated onto the new grid, to provide a good initial guess for the transport-reaction problem. Then the transport-reaction problem is solved. Based on the solution to the transport-reaction problem the electrode growth $\Delta L$ is calculated by solving \eqref{NonLin_PDE_impl_Euler} on the electrode boundary. At each mesh point a small random contribution $\Delta L_\mr{rand} = a^3 \Delta t J_\mr{rand}$ is then added to $\Delta L$. Finally, the new $x$ and $y$ positions are calculated by adding $n_x(\Delta L+ \Delta L_\mr{rand})$ and $n_y(\Delta L+ \Delta L_\mr{rand})$ to the old $x$ and $y$ positions.

The new $x$ and $y$ positions are exported to \textsc{MATLAB}. In \textsc{MATLAB} any inconsistencies arising from the electrode growth are resolved. If, for instance, the electrode surface intersects on itself, the points closest to each other at the intersection position are merged and any intermediate points are discarded. This corresponds to creating a hollow region in the electrode which is no longer in contact with the remaining electrolyte. The points are then interpolated so that they are evenly spaced, and exported to \textsc{COMSOL} so that the entire procedure can be repeated for a new time step.

The simulations are run on a standard work station with two 2.67 GHz Intel Xeon processors and 48 GB RAM. The electrodeposits shown in \secref{Results} typically take 2 days to run.

\subsection{Reduction of the computational domain}

\begin{figure}[!t]
    \includegraphics{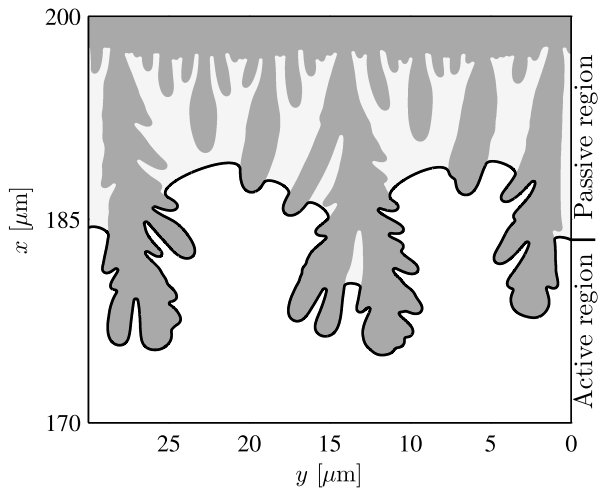}
    \caption{\figlab{Cut_line_plot} Example of the simplifying cutting procedure. The reduced interface (thick black line) divides the domain into an active region (white) and a passive region (light gray). The dark gray area shows the real cathode. The example is taken from a simulation with $c_0 = 1\ \SImM$ and $V_0 =10$ after deposition for 31 hours and 28 minutes. }
\end{figure}

At the cathode the mesh is much finer than in the remainder of the domain. The number of mesh points, and hence the computation time, therefore roughly scales with the length of the electrolyte-cathode interface. This has the unfortunate consequence that the computation time for each time step increases drastically, when branching structures emerge at the cathode. To lower the computation time we exploit the fact that the vast majority of the current enters near the tips of the dendritic structures. The parts of the cathode which are not near the tips can therefore be left fixed in time and thus removed from the simulation, without changing the results appreciably. This part of the domain is denoted the passive region. In regions where the current density is less than $0.001$ times the maximum value, we thus substitute the real, ramified electrode with a smooth line connecting the parts of the electrode with larger currents. The procedure is carried out in such a way that the real electrode surface can always be recovered from the reduced surface. For a few select examples we have verified that the results are unchanged by this simplifying procedure. In \figref{Cut_line_plot} is shown an example electrode surface together with the reduced surface. It is seen that the length of the electrolyte-cathode interface is heavily reduced by excluding parts of the electrode from the computation.

\subsection{Parameter values}

\begin{table}[!t]
\caption{\tablab{Parameters} Fixed parameter values used in the simulations.}
\begin{ruledtabular}
\begin{tabular}{lcl}

Parameter & Symbol & Value  \\ \hline

Cation diffusivity \cite{Lide2010} & $D_+$ & $0.714\times 10^{-9}\SIm^2/\SIs$ \rule{0mm}{2.5ex} \\

Anion diffusivity \cite{Lide2010} & $D_-$ & $1.065\times 10^{-9}\SIm^2/\SIs$ \\

Ion valence & $Z$ &  $2$ \\

Surface energy & $\gamma$ &$1.85\ \SIJ/\SIm^2$ \\

Temperature & $T$ & $300\ \SIK$ \\

Permittivity of water & $\epsw$ & $6.90\times 10^{-10}\SIF/\SIm$ \\

Charge-transfer coefficient & $\alpha$ & 0.5 \\

Reaction constant\footnote{Calculated using the exchange current $I_0=30\ \mr{A}/\SIm^2$ from Ref.~\cite{Turner1962} and $k_0 = I_0/(Ze)$.} & $k_0$ & $9.4\times 10^{19} \SIm^{-2}\SIs^{-1}$\\

Diameter of a copper atom\footnote{The cubic root of the volume per atom in solid copper \cite{Lide2010}.} & $a$ & $0.228\ \SInm$ \\

\end{tabular}
\end{ruledtabular}
\end{table}	

To limit the parameter space we choose fixed, physically reasonable values for the parameters listed in \tabref{Parameters}. The values are chosen to correspond to copper electrodes in a copper sulfate solution, see Ref.~\cite{Nielsen2015Arxiv} for details.

In Ref.~\cite{Nielsen2015Arxiv} we calculate the critical wavelength $\lambda_\mr{c}$, i.e. the smallest unstable perturbation  wavelength, for a range of parameters. We expect the critical wavelength to be the smallest feature in the problem, so we choose the mesh size accordingly. We set the mesh size at the electrode to $\Delta s = 0.1 \lambda_\mr{c}$, since our investigations, see \secref{Validation}, show that this is a suitable resolution. We also require that the mesh size does not exceed $0.1$ times the local radius of curvature. In the bulk part of the system we use a relatively coarse triangular mesh with mesh size $W/6$. Close to the cathode, in a region $l=0.5\ \SImum$ from the electrode surface, we use a triangular mesh with mesh size $l/4$. See \figref{Mesh_plot2} for a meshing example.

\begin{figure}[!t]
    \includegraphics[width=0.8\columnwidth]{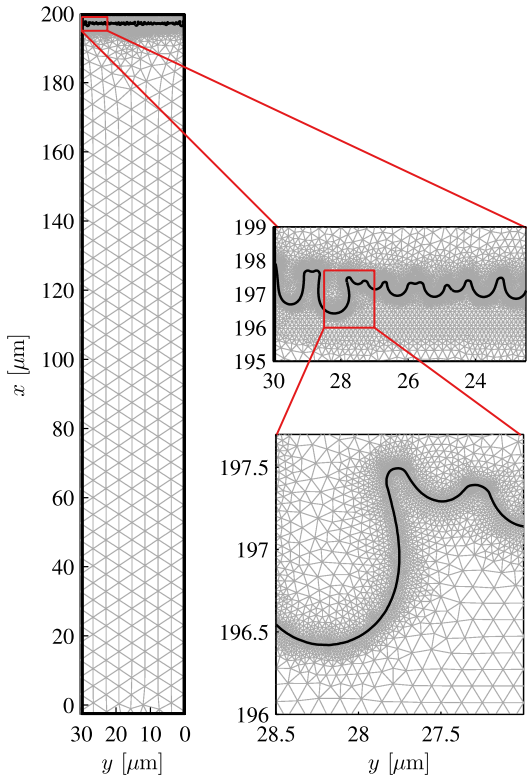}
    \caption{\figlab{Mesh_plot2} (Color online) Example of domain meshing at varying magnification. The example is taken from a simulation with $c_0 = 1\ \SImM$ and $V_0 =10$ after deposition for 7 hours and 50 minutes. The wiggly black line is the cathode surface. The light gray lines are the mesh boundaries and the dark (red) lines show the sections that are magnified. The mesh elements above the cathode surface are only used for storing the solution between time steps. }
\end{figure}

We choose a fixed value for the bin depth $\Delta h = 0.2 \lambda_\mr{c}$. In accordance with the analysis in \appref{Init_growth} the time step $\Delta t$ is chosen so that it is always smaller than $0.5/\Gamma_\mr{max}$. In addition, the time step is chosen so that at each point on the cathode, the growth during the time step is smaller than the local radius of curvature.

We fix the length $L$ to $100 \ \SImum$. According to the time-scale analysis in \secref{Sol_method} and the instability growth rates found in Ref.~\cite{Nielsen2015Arxiv}, the quasi-steady state approximation is valid for $L=100\ \SImum$. The width $W$ of the system is set to $W=200 \lambda_\mr{c}$, rounded to the nearest micrometer. This makes for a system that is broad enough to exhibit interesting growth patterns, while having a reasonable computation time. The growth is somewhat affected by the symmetry boundaries at $y=0$ and $y=W$, especially at later times.

These choices leave us with two free parameters, which are the bias voltage $V_0$ and the electrolyte concentration $c_0$. We solve the system for $c_0 = \{1\ \SImM,10\ \SImM,100\ \SImM\}$ and $V_0 = \{10,20,30\}$.

\subsection{Validation}  \seclab{Validation}

\begin{figure}[!t]
    \includegraphics[width=0.9\columnwidth]{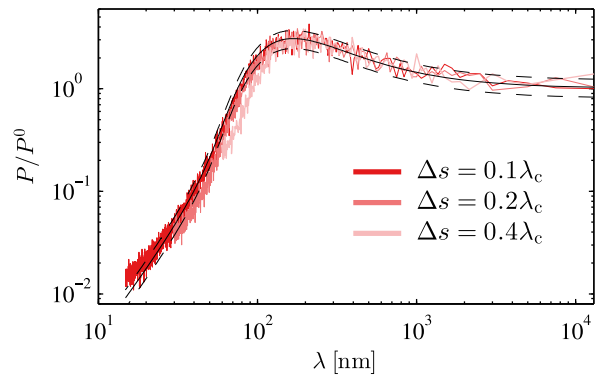}
    \caption{\figlab{Power_spectrum} (Color online) Power spectra averaged over 50 runs for three different mesh sizes, $\Delta s=\{ 0.1\lamc, 0.2\lamc, 0.4\lamc \}$. In each run we used $M=100$ time steps of $\Delta t = 0.64 \ \SIs$ and the parameter values $c_0=10 \ \SImM$, $L= 100 \ \SImum$, and $V_0=30$. The full black line shows the analytical result and the dashed black lines show the analytical standard error on the mean. The result for $\Delta s =0.1 \lamc$ is shown in dark (red), the result for $\Delta s =0.2 \lamc$ is shown in medium (red), and the result for $\Delta s =0.4\lamc$ is shown in bright (red). }
\end{figure}

The random nature of the phenomena we are investigating poses obvious challenges when it comes to validating the numerical simulations. The individual steps in the computation can be, and have been, thoroughly tested and validated, but testing whether the aggregate behavior after many time steps is correct is a much taller order. At some level, we simply have to trust that, if the individual steps are working correctly, then the aggregate behavior is also correct. To support this view, there is one test we can make of the aggregate behavior in the very earliest part of the simulation.

In the early stages of the simulation the electrode surface is deformed so little, that the linear stability analysis from \cite{Nielsen2015Arxiv} should still be valid. We thus have an analytical expression for the wavelength dependent growth rate $\Gamma$, which we can compare with the growth rates found in the numerical simulations. In \appref{Init_growth} we calculate an expression for the average power spectrum of the cathode interface after deposition for a time $t_\mr{tot}$, given the type of noise described in \secref{Noise},
\begin{align} \eqlab{avg_Pn}
\langle P_n \rangle &=a^6 \frac{J_+ }{2 \Delta h W \Gamma_n}\left [ e^{2\Gamma_n  t_\mr{tot}} -1 \right ],
\end{align}
where $\Gamma_n$ is the growth rate of the $n$'th wavelength $\lambda_n =W/n$ component in the noise spectrum. We also find the standard deviation $\mr{SD}(P_n)$ of the power spectrum
\begin{align}
\mr{SD}(P_n) \approx \sqrt{2} \langle P_n \rangle.
\end{align}
Because the standard deviation of $P_n$ is so large compared to the mean value, it is necessary to average over many runs before a meaningful comparison with \eqref{avg_Pn} can be made. Averaging the power spectrum over 50 simulations brings the standard error on the mean down to 20 percent times the mean value, at which point a reasonable comparison can be made. In \figref{Power_spectrum} the power spectrum averaged over 50 runs is shown for three different mesh sizes, $\Delta s= \{ 0.1\lamc, 0.2\lamc, 0.4\lamc \}$. In each run we used $M=100$ time steps of $\Delta t = 0.64 \ \SIs$ and the parameter values $c_0=10 \ \SImM$, $L= 100 \ \SImum$, and $V_0=30$. The chosen step size corresponds to $0.01/\Gammax$. The analytical result \eqrefNoeq{avg_Pn} is also shown together with the standard error on the mean. The power spectra are normalized with the power $P^0$ obtained for $\Gamma =0$,
\begin{align}
P^0 = a^6 \frac{J_+  t_\mr{tot}}{\Delta h W}.
\end{align}

It is seen that for $\Delta s = 0.4 \lamc$ some of the power in the small wavelength components is filtered out. As the mesh size is decreased to $\Delta s = 0.2 \lamc$ and $\Delta s =0.1 \lamc$ the low wavelength components are represented increasingly well.

In the above treatment, the time step was chosen very small compared to the instability time scale, $\Delta t = 0.01/\Gammax$. This was done to approach the limit of continuous time, and thus enable the best possible comparison with the analytical theory. Such a short time step is, however, impractical for the much longer simulations in the remainder of the paper. In those simulations we use time steps as large as $\Delta t = 0.5/\Gammax$. Due to the coarser time resolution employed in the remaining simulations, we expect their power spectrum to deviate somewhat from the almost ideal behavior seen in \figref{Power_spectrum}.

\section{Results}
\seclab{Results}

\begin{figure*}[!t]
    \includegraphics[width=17 cm]{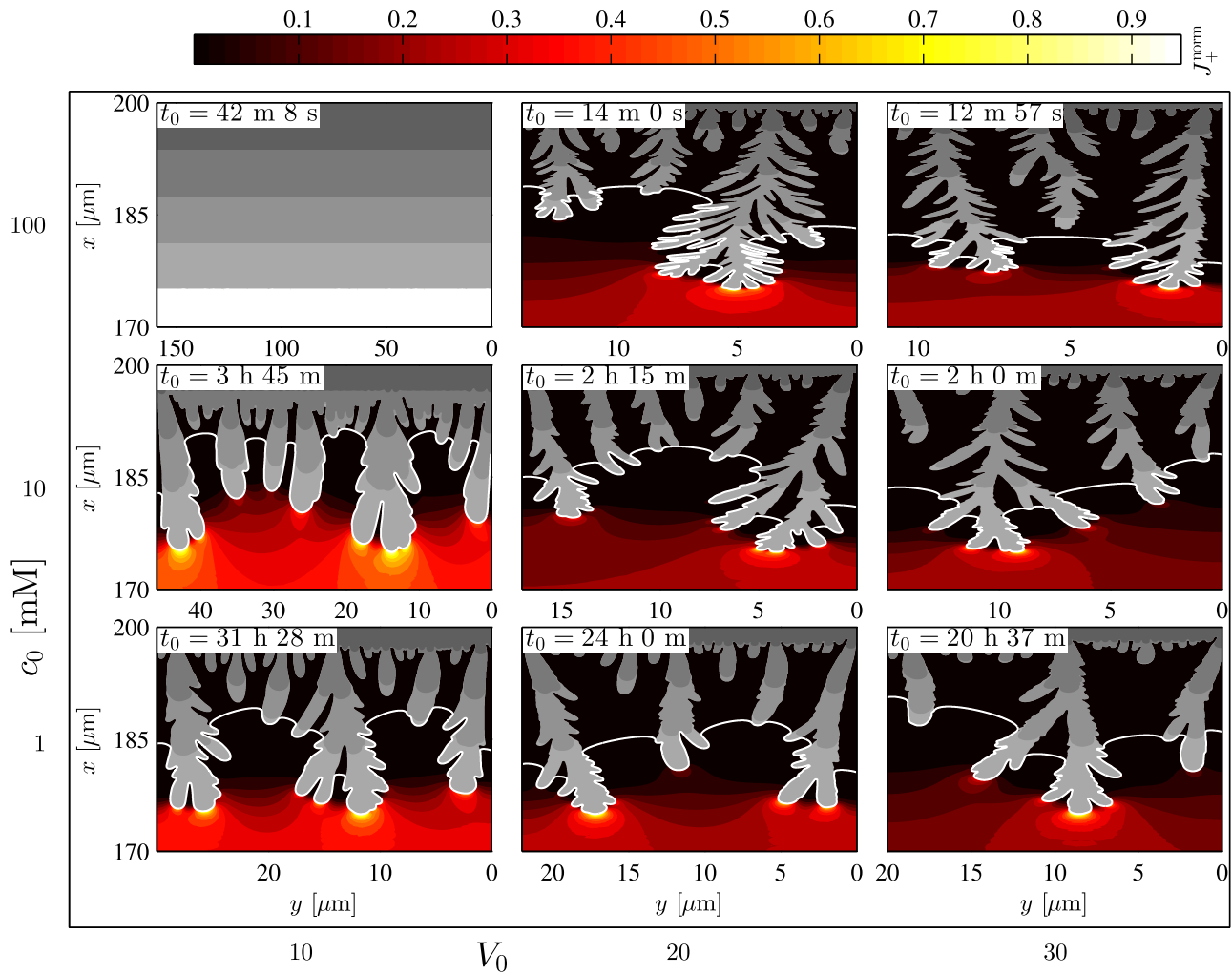}
    \caption{\figlab{Several_dendrites_with_fields} (Color online) Electrodeposits in the $V_0$-$c_0$ plane obtained for $L=100~\SImum$, $c_0 = \{1\ \SImM,10\ \SImM,100\ \SImM\}$ and $V_0 = \{10,20,30\}$. The aspect ratio varies between the panels, since the width $W$ of the simulated region is always set to $200 \lambda_\mr{max}$. The gray area has different shades corresponding to times $t_0$ (light), $0.75t_0$ (darker), $0.5t_0$ (darker yet), $0.25t_0$ (darkest). The white line indicates the reduced surface at time $t_0$. The contours in the liquid represent the relative magnitude of the cation current.}
\end{figure*}

We let the simulations run until the cathode has grown $25\ \SImum$. The time $t_0$ it takes to reach this point varies greatly with the parameters, mainly because the limiting current scales with $c_0$. In \figref{Several_dendrites_with_fields} the cathode surfaces are shown along with heat plots showing the relative magnitude of the current density at the last time step. The white line shows the position of the reduced interface at the last time step, and the gray area shows the actual position and shape of the cathode. The gray electrodeposits have different shades corresponding to $0.25t_0$, $0.5t_0$, $0.75t_0$, and $t_0$. The heat plot shows the value of $J_+^{\mr{norm}}$, which is the magnitude of the cation current density normalized with its maximum value. In each panel $J_+^{\mr{norm}}$ thus varies from 0 to 1.

To investigate the reproducibility of the results we have repeated the simulation of the $c_0=1\ \SImM$, $V_0=10$ system two times. All three electrodeposits are seen in \figref{Triplicate_plot}. The electrodeposits are clearly different from one another, as expected for a random process, but they are also seen to share some general features. These shared features are most easily appreciated by comparing the electrodeposits in  \figref{Triplicate_plot} to the electrodeposits in \figref{Several_dendrites_with_fields}. It is seen that the electrodeposits in \figref{Triplicate_plot} are much more similar to each other, than to any of the remaining electrodeposits in \figref{Several_dendrites_with_fields}. Thus, the results are reproducible in the sense, that the random electrodeposits have some general features that are determined by the parameter values.

\begin{figure}[!t]
    \includegraphics{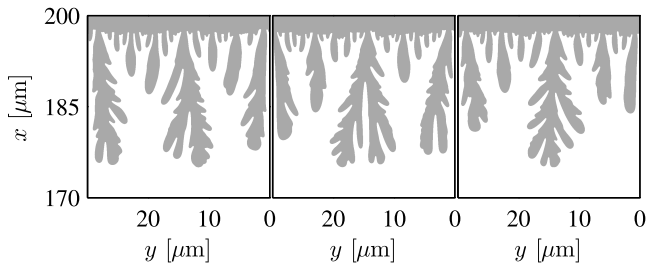}
    \caption{\figlab{Triplicate_plot} Three simulations of electrodeposits using the same parameter values $L=100 \ \SImum$, $c_0 = 1\ \SImM$, and $V_0=10$. The electrodeposits are clearly different from one another, but they do share some general features. }
\end{figure}

When interpreting the plots in \figref{Several_dendrites_with_fields}, we should be mindful that the aspect ratio is not the same in each panel. The reason for this is that the vertical axis has the same length, $30\ \SImum$, in each panel, while the length of the horizontal axis, $W$, varies between panels. In \figref{Dendrite_carpet_plots} we show adapted versions of the panels from \figref{Several_dendrites_with_fields}. The subfigures in \figref{Dendrite_carpet_plots} are created by repeatedly mirroring the subfigures from \figref{Several_dendrites_with_fields} until their horizontal length is $100\ \SImum$. Obviously, the resulting extended cathodes are somewhat artificial, since we have imposed some symmetries, which would not be present in a simulation of a system with $W=100 \ \SImum$. Nevertheless, we find the subfigures in \figref{Dendrite_carpet_plots} useful, since they give a rough impression of the appearance of wider systems and allow for easier comparison of length scales between panels.

\subsection{Rationalizing the cathode morphologies}

The cathode morphologies observed in \figref{Several_dendrites_with_fields} and \figref{Dendrite_carpet_plots} are a function of several factors, some of which we attempt to outline below. First, we consider the time $t_0$ it takes before part of the cathode reaches $x=175\ \SImum$. As seen from \eqref{Simple_deltaL}, this time is mainly a function of the limiting current. This explains the approximately inverse scaling with $c_0$. The current density also increases with $V_0$, which is why the time $t_0$ decreases slightly as $V_0$ increases. Finally, the time $t_0$ scales with the filling factor. This is the reason why $t_0$ is much larger in the upper left panel of \figref{Several_dendrites_with_fields}, than in either of the two other top row panels.

It is apparent from the lack of ramified growth, that the cathode in the upper left panel in \figref{Several_dendrites_with_fields} is considerably more stable than the other systems in the leftmost column. To explain this variation in stability, we refer to Fig.~6 in Ref.~\cite{Nielsen2015Arxiv}. There it is shown that the instability length scale is on the order of $50\ \SImum$ for $c_0 = 100\ \SImM$ at $V_0=10$, while it is considerably lower for $c_0 = 10\ \SImM$ and $c_0 = 1\ \SImM$. Fig.~6 in Ref.~\cite{Nielsen2015Arxiv} also shows that for $V_0 >18$ the instability length scale decreases in size as the concentration increases. The same tendency is observed in \figref{Dendrite_carpet_plots}.

From the subfigures in \figref{Dendrite_carpet_plots} it appears that there is a connection between the thickness of the layer deposited before the instabilities develop, and the characteristic length scale of the ramified electrodeposits. The analysis in Ref.~\cite{Nielsen2015Arxiv} suggests that there is indeed such a connection and, moreover, that both lengths should scale with the most unstable wavelength for the given parameters. To test this assertion, we plot the thickness $\delta_\mathrm{inst}$ of the layer deposited before the instabilities develop, versus the most unstable wavelength $\lambda_\mr{max}$. We exclude the $c_0 = 100\ \SImM$, $V_0=10$ system, since instabilities have not yet developed in this system. The resulting plot is seen in \figref{Instability_dimension}(a) together with a linear fit. Although there is a good amount of scatter around the linear fit, it is seen to capture the general trend reasonably well.

We would like to make a similar plot with the characteristic length scale $\delta_\mr{char}$ of the ramified electrodeposits on the $y$-axis. To extract $\delta_\mr{char}$, we follow the approach in Ref.~\cite{Genau2013} and calculate the so-called Minkowski dimension of each electrodeposit. In doing this we only consider the part of the electrodeposit lying between $170\ \SImum$ and $190\ \SImum$, and as before we exclude the $c_0 = 100\ \SImM$, $V_0=10$ system. In this work we are actually not interested in the Minkowski dimension itself, but rather in a partial result that follows from the analysis. In a range of length scales the electrodeposits appear roughly fractal, but below a certain length scale the electrodeposits are locally smooth. The length scale at which this transition occurs can be extracted from the analysis, and we use this length as the characteristic length scale $\delta_\mr{char}$ of the electrodeposit, see \appref{Minkowski_dimension}. In \figref{Instability_dimension}(b) we plot $\delta_\mr{char}$ versus $\lammax$. Also here, we find a roughly linear behavior.

\begin{figure*}[!t]
    \includegraphics[width=17 cm]{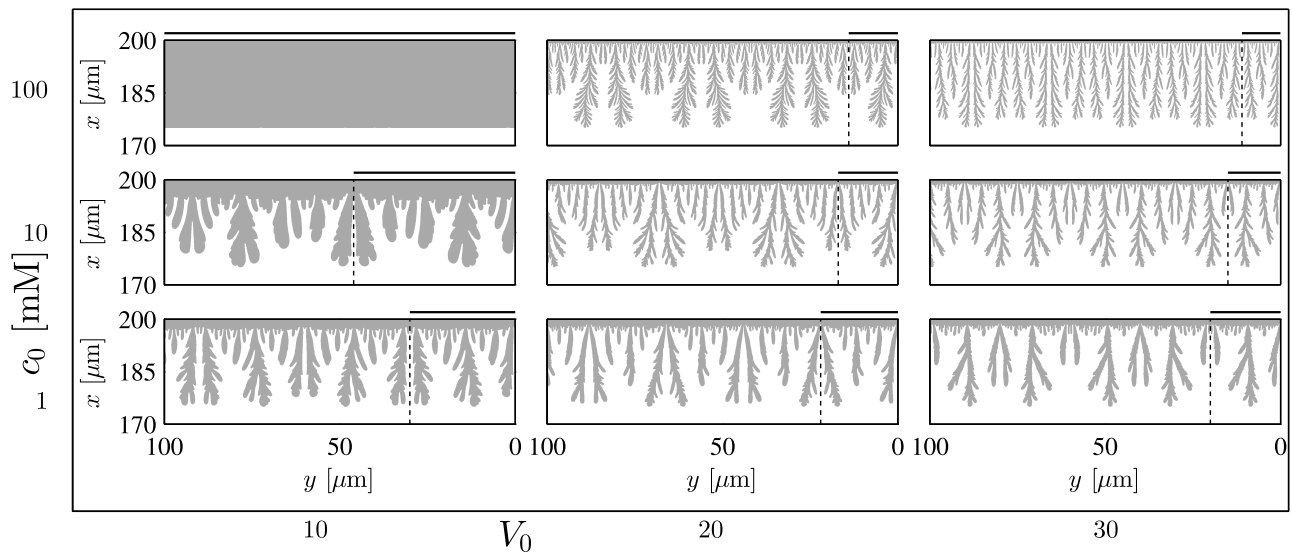}
    \caption{\figlab{Dendrite_carpet_plots} Extended electrodeposits in the $V_0$-$c_0$ plane obtained by mirroring those from \figref{Several_dendrites_with_fields} in their symmetry axes until the width equals $100\ \SImum$. The dashed line indicates the first mirror plane, i.e. the part between the dashed line and $y=100 \ \SImum$ are obtained by repeating the part marked by a black line.  }
\end{figure*}

\begin{figure}[!t]
    \includegraphics[width=0.9\columnwidth]{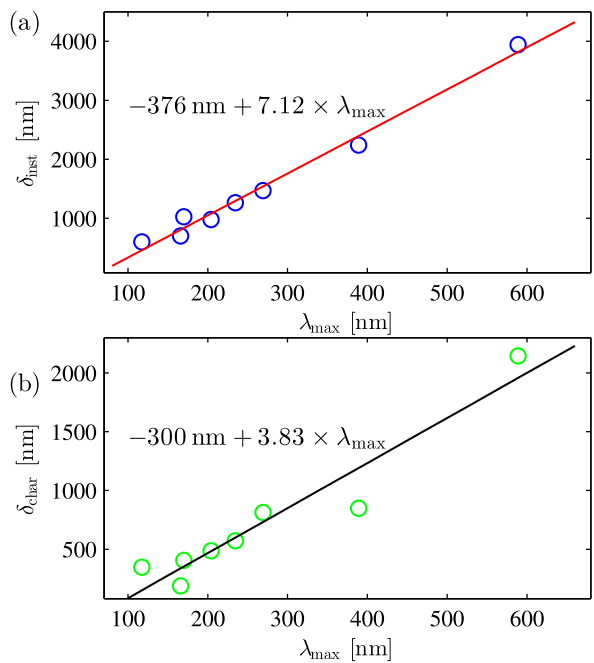}
    \caption{\figlab{Instability_dimension} (Color online) (a): The instability length scale $\delta_\mr{inst}$ obtained from the simulations, plotted versus the most unstable wavelength $\lambda_\mr{max}$. Also, a linear fit highlighting the roughly linear dependence is shown. (b): The characteristic length scale $\delta_\mr{char}$ obtained from the simulations, plotted versus the most unstable wavelength $\lambda_\mr{max}$. Also, a linear fit highlighting the roughly linear dependence is shown.    }
\end{figure}

Evidently, $\lammax$ plays an important role for the morphology of the electrodeposits. However, $\delta_\mr{inst}$ and $\delta_\mr{char}$ alone are not sufficient to characterize the electrodeposits. As seen in the top row of \figref{Dendrite_carpet_plots}, the characteristic length scale $\delta_\mr{char}$ varies very little between $V_0=20$ and $V_0=30$. Yet, the morphology still changes appreciably. The reason for this change in morphology is probably that the gradient in electrochemical potential increases near the cathode as the bias voltage is increased. The larger the electrochemical gradient is, the more the system will favor deposition at the most protruding parts of the electrodeposits. For large voltages we therefore expect long and narrow electrodeposits, whereas we expect dense branching electrodeposits for low voltages.

\section{Discussion}

Our model improves on existing models in three important ways: it can treat systems at overlimiting current including the extended space-charge region, it allows for a proper reaction boundary condition, and it can be tested against results from sharp-interface stability analyses. Our model is, however, not without issues of its own. Perhaps the most apparent of these is the quasi-steady-state assumption. This assumption limits the applicability of the model to short systems, in which the diffusion time is small compared to the deposition time, as discussed in \secref{Sol_method}. In principle the phase-field models are superior to our model in this aspect, since they do not have this limitation. However, it is not of practical relevance, as all of the published phase-field simulations are for systems so short that the quasi-steady-state assumption is valid anyway \cite{Shibuta2007, Liang2014, Cogswell2015}.

It is well known, that the strong electric fields at the dendrite tips give rise to electroosmotic velocity fields in the system \cite{Fleury1992, Fleury1993, Huth1995}. To simplify the treatment and bring out the essential physics of electrodeposition, we have chosen not to include fluid dynamics and advection in our model. However, it is straightforward to include these effects, see for instance our previous work \cite{Nielsen2014b}.

One of the main advantages the sharp-interface model has over the phase-field models, is that it allows for the implementation of proper reaction boundary conditions. The standard Butler--Volmer model used in this paper is a first step towards realistic reaction boundary conditions. As elaborated by Bazant in Ref.~\cite{Bazant2013}, there are other reaction models, such as Marcus kinetics, which might better describe the electrode reactions. Also, the standard Butler--Volmer model has the contentious assumption that the overpotential is the total potential drop over both the electrode-electrolyte interface and the Debye layer. A more realistic approach might be to model the Debye layer explicitly or include the Frumkin correction to the Butler--Volmer model \cite{Soestbergen2012}. Furthermore, a proper reaction expression should take the crystal structure of the material into account. There are simple ways of implementing crystal anisotropy in the surface tension term, see for instance Refs.~\cite{Cogswell2015,Kobayashi1993}, but again, to keep the model simple we have chosen not to include anisotropy at the present stage. Any of the above mentioned reaction models can be easily implemented in the framework of the sharp-interface model, and as such the specific Butler--Volmer model used in this work does not constitute a fundamental limitation.

More broadly, our sharp-interface model includes, or allows for the easy inclusion of, most effects that are important for electrodeposition in 2D. A natural next step is therefore to see how our results compare to experimental electrodeposits. Unfortunately, most such experimental data are viewed at the millimeter or centimeter scale, whereas our simulation results are at the micrometer scale. In one paper, Ref.~\cite{Lin2011}, the electrodeposits are probed at the micrometer scale, but the results do not make for the best comparison, since the morphology of their electrodeposits was a result of adding a surface active molecule. We hope that as more experimental results become available, it will be possible to perform rigorous tests of our model.

%
%
%
%
%
%

%
%
%
%
%
%

\section{Conclusion}

We have developed a sharp-interface model of electrodeposition, which improves on existing models in a number of ways. Unlike earlier models, our model is able to handle sharp-interface boundary conditions, like the Butler-Volmer boundary condition, and it readily deals with regions with non-zero space-charge densities. A further advantage is that our model handles the physical problem in much the same way as done in various linear stability analyses. We can thus obtain a partial validation of our model by comparing its predictions with those of a linear stability analysis. As of now, the main weakness of our model is that it assumes quasi-steady state in the transport equations. For the systems studied in this paper this is a reasonable assumption, since the diffusion time is small compared to the instability time. In future work we want to extend the model to the transient regime, so that larger systems can be treated as well.

The main aim of this paper has been to establish the sharp-interface method, but we have also included a study of the simulated electrodeposits. An interesting observation is, that the characteristic length scale of the electrodeposits seems to vary linearly with the size of the most unstable wavelength. This exemplifies a promising application of our sharp-interface model, namely as a tool to develop a more quantitative understanding of electrodeposits and their morphology.

\acknowledgements

We thank Edwin Khoo and Prof. Martin Z. Bazant for valuable discussions of electrode reactions and the growth mechanisms.

\appendix

\section{Initial growth}  \chaplab{Init_growth}

In the initial part of the simulation the electrode is so flat that the linear stability analysis from Ref.~\cite{Nielsen2015Arxiv} gives a good description of the growth. We parameterize the cathode position as
\begin{align}
x= X(t) + f(y,t),
\end{align}
where $f(y,t)$ is the $y$-dependent deviation from the mean electrode position $X(t)$. According to the linear stability analysis each mode grows exponentially in time with the growth factor $\Gamma$. After a time $t$ an initial perturbation,
\begin{align}  \eqlab{an_f0}
f(y,0)  = \sum_{n=1}^{N} a_n e^{i k_ny},
\end{align}
has therefore evolved to
\begin{align}
f(y,t)  = \sum_{n=1}^{N} a_n  e^{\Gamma_n t} e^{i k_ny}.
\end{align}
We note that some of the growth rates $\Gamma_n$ can be negative. In our simulation we add new perturbations with small time intervals, which we, for the purpose of this analysis, assume to be evenly spaced. After $M$ time intervals $\Delta t$ the surface is therefore described by
\begin{align}
 f(y,M\Delta t) = \sum_{m=0}^M\sum_{n=1}^{N} a_{nm} e^{\Gamma_n (M-m)\Delta t}e^{i k_ny}.
 \end{align}
We are interested in the average power of each mode
\begin{align}
\langle P_n\rangle =\left \langle \left |\sum_{m=0}^M a_{nm} e^{\Gamma_n (M-m)\Delta t} \right |^2 \right \rangle.
\end{align}
The coefficients are random and uncorrelated with zero mean. On average the cross-terms in the sum therefore cancel and we can simplify,
\begin{align}
\langle P_n\rangle& =\left \langle \sum_{m=0}^M |a_{nm}|^2 e^{2\Gamma_n (M-m)\Delta t}  \right \rangle \nn \\
&=\langle|a_{n}|^2\rangle \sum_{m=0}^M   e^{2\Gamma_n (M-m)\Delta t} \nn \\
&=\langle|a_n|^2\rangle \frac{e^{2\Gamma_n (M+1)\Delta t} -1}{e^{2\Gamma_n\Delta t} -1}.    \eqlab{anl_Power_spectrum}
\end{align}
The variance of the power is given as
\begin{align}
\mr{Var}(P_n) =  \langle P_n^2\rangle-\langle P_n\rangle^2.
\end{align}
The first of these terms is
\begin{align}
\langle P_n^2\rangle & =\left \langle \left (\left |\sum_{m=0}^M a_{nm} e^{\Gamma_n (M-m)\Delta t} \right |^2 \right )^2\right \rangle \nn \\
&=e^{4\Gamma_n M\Delta t} \left \langle \left (\left |\sum_{m=0}^M a_{nm} q^{m} \right |^2 \right )^2\right \rangle,
\end{align}
where $q=e^{-\Gamma_n \Delta t}$. Writing out the absolute value
\begin{align}
\langle P_n^2\rangle & =e^{4\Gamma_n M\Delta t} \left \langle \left ( \sum_{m'=0}^M \sum_{m=0}^M a_{nm}a_{nm'}^* q^{m+m'}  \right )^2\right \rangle,
\end{align}
where superscript $*$ denotes complex conjugation. Because the coefficients are uncorrelated with mean 0, only the terms including $|a_{nm}|^2|a_{nm'}|^2$ survive in the average of the square,
\begin{align}
\langle P_n^2\rangle & =e^{4\Gamma_n M\Delta t} \left \langle  \frac{1}{2}\sum_{m'=0}^M \sum_{m=0}^M 6|a_{nm}|^2|a_{nm'}|^2 q^{2(m+m')}  \right \rangle \nn \\
& =3e^{4\Gamma_n M\Delta t} \sum_{m'=0}^M \sum_{m=0}^M \left \langle  |a_{nm}|^2|a_{nm'}|^2\right \rangle q^{2(m+m')}.
\end{align}
Here, the factor of six comes from the binomial coefficient and the factor of a half takes into account that the double sum counts each combination twice.
Now, there are two possibilities; either $m\neq m'$ or $m= m'$. In the first case $|a_{nm}|^2$ and $|a_{nm'}|^2$ are uncorrelated, meaning that
\begin{align}
\left \langle  |a_{nm}|^2|a_{nm'}|^2\right \rangle =\left \langle  |a_{n}|^2\right \rangle^2.
\end{align}
Whereas if $m= m'$, then
\begin{align}
\left \langle  |a_{nm}|^2|a_{nm'}|^2\right \rangle =\left \langle  |a_{n}|^4\right \rangle.
\end{align}
This means that
\begin{align}
\langle P_n^2\rangle & =3e^{4\Gamma_n M\Delta t} \left \langle  |a_{n}|^2\right \rangle^2 \sum_{m'\neq m}^M \sum_{m=0}^M  q^{2(m+m')} \nn \\
&\quad + 3e^{4\Gamma_n M\Delta t} \left \langle  |a_{n}|^4\right \rangle  \sum_{m=0}^M  q^{4m} \nn \\
&= 3e^{4\Gamma_n M\Delta t} \left \langle  |a_{n}|^2\right \rangle^2 \sum_{m'=0}^M \sum_{m=0}^M  q^{2(m+m')} \nn \\
&\quad + 3e^{4\Gamma_n M\Delta t} \left (\left \langle  |a_{n}|^4\right\rangle -\left \langle  |a_{n}|^2\right \rangle^2\right )  \sum_{m=0}^M  q^{4m} \nn \\
&=3\langle P_n \rangle^2 \nn \\
&\quad + 3e^{4\Gamma_n M\Delta t} \left (\left \langle  |a_{n}|^4\right\rangle -\left \langle  |a_{n}|^2\right \rangle^2\right )  \frac{q^{4(M+1)}-1}{q^4-1}.
\end{align}
The variance of the power is thus given as
\begin{align}
\mr{Var}(P_n) =2\langle P_n \rangle^2+\left (\left \langle  |a_{n}|^4\right\rangle -\left \langle  |a_{n}|^2\right \rangle^2\right )  \frac{e^{4\Gamma_n (M+1)\Delta t}-1}{e^{4\Gamma_n \Delta t}-1}.
\end{align}
If $\Gamma_n \Delta t \ll 1$ we can expand the denominators of $\langle P_n \rangle^2$ and the last term. We find that they scale as $4 (\Gamma_n \Delta t)^2$ and $4\Gamma_n \Delta t$, respectively. In the limit $\Gamma_n \Delta t \ll 1$ the first term thus dominates over the second, so to a good approximation we have
\begin{align}
\mr{Var}(P_n) \approx 2\langle P_n \rangle^2, \\
\mr{SD}(P_n) \approx \sqrt{2} \langle P_n \rangle.
\end{align}

In the simulations the surface perturbations have the form
\begin{align} \eqlab{bn_f0}
f(y,0) = \sum_{n=1}^N b_n h(y-n\Delta y),
\end{align}
where,
\begin{align}
h(y) = \left\{\begin{array}{cl} 1, & 0 \leq y \leq \Delta s,\\
0, & \text{else}.
\end{array} \right.
\end{align}
We take the absolute square of $f(y,0)$ given as both \eqref{an_f0} and \eqref{bn_f0}, and integrate over the domain to obtain
\begin{align}
\int_0^W |f(y,0)|^2\ \mr{d}y  &= \sum_{n=1}^N |b_n|^2\int_0^W | h(y-n\Delta y)|^2\ \mr{d}y \nn \\
&= \Delta s\sum_{n=1}^N |b_n|^2, \\
\int_0^W |f(y,0)|^2\ \mr{d}y  &= \sum_{n=1}^N |a_n|^2\int_0^W | e^{i k_ny}|^2\ \mr{d}y \nn \\
&= W \sum_{n=1}^N |a_n|^2.
\end{align}
The mean square of $b_n$ is thus related to the mean square of $a_n$ as
\begin{align}
\langle |a_n|^2 \rangle = \frac{\Delta s}{W}\langle |b_n|^2 \rangle=\frac{1}{N}\langle |b_n|^2 \rangle.
\end{align}
From \eqref{DL_rand} we have that
\begin{align}
\langle |b_n|^2 \rangle = a^6 \frac{J_+ \Delta t}{\Delta h\Delta s} .
\end{align}
Inserting in \eqref{anl_Power_spectrum} we find
\begin{align}
\langle P_n \rangle& = \frac{1}{N} a^6 \frac{J_+ \Delta t}{\Delta h\Delta s} \frac{e^{2\Gamma_n (M+1)\Delta t} -1}{e^{2\Gamma_n\Delta t} -1} \\
&=a^6 \frac{J_+ \Delta t}{\Delta h W} \frac{e^{2\Gamma_n (t_\mr{tot}+\Delta t)} -1}{e^{2\Gamma_n\Delta t} -1},
\end{align}
which is seen to be independent of the bin size $\Delta s$. We also introduced the total time $t_\mr{tot} = M \Delta t$. In a consistent scheme the power spectrum should of course only depend on the total time, and not on the size $\Delta t$ of the time steps. For small values of $\Gamma_n\Delta t$ we can expand the denominator and neglect the $\Delta t$ in the nominator,
\begin{align}
\langle P_n \rangle& \approx a^6 \frac{J_+ }{2 \Delta h W \Gamma_n}\left [ e^{2\Gamma_n  t_\mr{tot}} -1 \right ].
\end{align}
So, as long as $2\Gamma_n\Delta t \ll 1$ the power spectrum does not depend on the size of the time step.

For larger values of $2\Gamma_n\Delta t$ the power spectrum does depend on the size of the time step. However, as long as $2\Gamma_n\Delta t \lesssim 1$, we do not expect the overall morphology of the electrode to have a significant dependence on the time step.

\section{Characteristic length scale}  \chaplab{Minkowski_dimension}

\begin{figure}[!t]
    \includegraphics{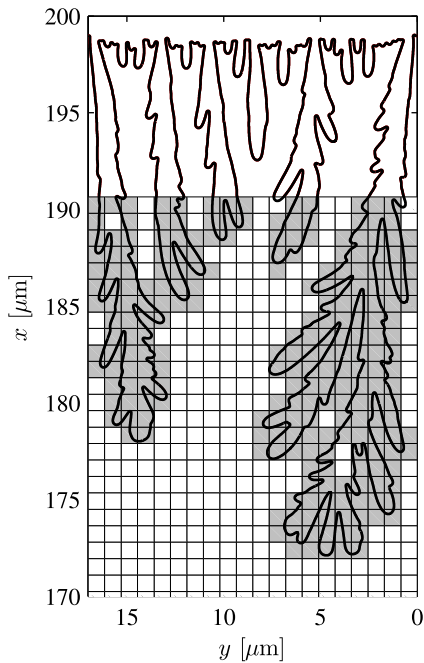}
    \caption{\figlab{Minkowski_plot} The box-counting method illustrated on the electrodeposit obtained for $c_0=10\ \SImM$ and $V_0=20$. The boxes that cover part of the deposit perimeter are shown in gray and the remaining boxes are shown in white. In this example the grid size is $\epsilon = 0.85 \ \SImum$ and the number of boxes it takes to cover the perimeter is $N(\epsilon) =234$. }
\end{figure}

\begin{figure}[!t]
    \includegraphics{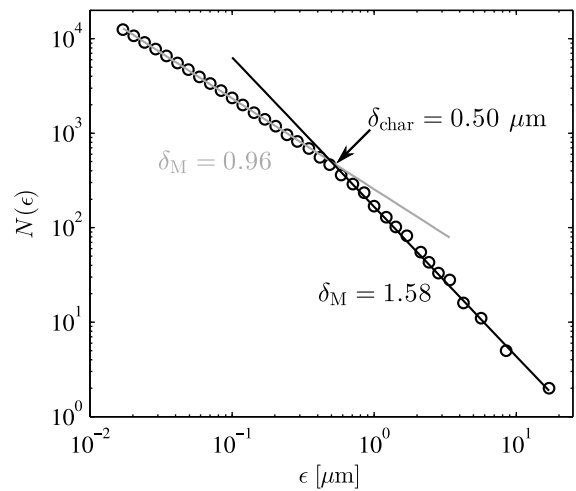}
    \caption{\figlab{Characteristic_length}   The number $N(\epsilon)$ of boxes it takes to cover the electrodeposit plotted vs the box side length $\epsilon$. A linear fit is shown in each of the two approximately linear regions, and the Minkowski dimension in each region is indicated. The crossing point between the linear fits is marked by an arrow, and the characteristic dimensions $\delta_\mr{char}=0.50 \ \SImum$ is calculated based on this crossing point.    }
\end{figure}

To find the characteristic length scale $\delta_\mr{char}$ of the ramified electrodeposits we follow Ref.~\cite{Genau2013} and use the box-counting method to calculate the Minkowski dimension of the deposits. We place a square grid with side length $\epsilon$ over each deposit, and count the number $N(\epsilon)$ of boxes it takes to completely cover the perimeter of the part of the deposit lying between $x=170\ \SImum$ and $x=190\ \SImum$. An example is shown in \figref{Minkowski_plot}.

For a proper fractal geometry, the Minkowski dimension is defined as
\begin{align}
\delta_\mr{M} =- \lim_{\epsilon \rightarrow 0} \frac{\ln\big[N(\epsilon)\big]}{\ln(\epsilon)}.
\end{align}
The electrodeposits we are investigating are not fractal at all length scales, but in a range of length scales, we can calculate an approximate Minkowski dimension as the negative slope in a $\ln\big[N(\epsilon)\big]$ vs $\ln(\epsilon)$ plot. In \figref{Characteristic_length} such a plot is seen, together with linear fits in each of the two approximately linear regions. The Minkowski dimension at small $\epsilon$ is nearly unity, indicating that the deposit perimeter is locally smooth at this length scale. For larger values of $\epsilon$ the Minkowski dimension deviates from unity, because the deposit is approximately fractal in this size range. At the transition point between these two regions is the smallest length scale, which is related to the morphology of the electrodeposit. This length scale we denote the characteristic length $\delta_\mr{char}$. Technically, we define $\delta_\mr{char}$ as the point where the linear fits from each region cross each other, as indicated in \figref{Characteristic_length}.

\newpage


%

\end{document}